\documentclass[aps,preprintnumbers,amsmath,amssymb,nofootinbib]{revtex4}

\usepackage{float}
\usepackage{graphics,epsfig}
\usepackage{bm}
\usepackage{slashed}
\usepackage{graphicx}%Include figure files
\usepackage{amsmath}
\usepackage{amsfonts}
\usepackage{amssymb}
\usepackage{color}%
\usepackage{dcolumn}% Align table columns on decimal point
\usepackage{hyperref}
\usepackage{enumerate}
\providecommand{\U}[1]{\protect\rule{.1in}{.1in}}

\begin{document}
\baselineskip=0.6 cm \title{Holographic heat engine with momentum relaxation}

\author{Li-Qing Fang$^{1}$}
\email{fangliqing@outlook.com}
\affiliation{
$^{1}$School of Physics and Electronic Information, Shangrao Normal University, Shangrao 334000, China.}
\author{Xiao-Mei Kuang$^{2}$}
\email{xmeikuang@gmail.com}
\affiliation{$^{2}$Center for Gravitation and Cosmology, College of Physical Science
and Technology, Yangzhou University, Yangzhou 225009, China}
\vspace*{0.2cm}
\begin{abstract}
\baselineskip=0.6 cm
\begin{center}
{\bf Abstract}
\end{center}
We investigate the  heat engine defined via black hole with momentum relaxation, which is introduced by massless axion fields. We first study the extended thermodynamical properties of the black hole and then apply it to define a heat engine. Then, we analyze how the momentum relaxation affects the  efficiency of the heat engine in the limit of high temperature.  We find that depending on the schemes of specified parameters in the engine circle,  the influence of momentum relaxation on the efficiency of the heat engine behaves novelly, and the qualitative behaviors do depend on the dimension of the gravity theory.
\end{abstract}

%\pacs{11.25.Tq, 04.50.Gh, 71.10.-w}
\maketitle
\newpage
\vspace*{0.2cm}

\section{Introduction}
%%%%%%%%%%%%%%%%%%%%%%%%%%%%%%%%
 Thermodynamics of the anti de Sitter (AdS) black holes have been attracting more and more attention in the development of holographic gauge/gravity duality, because they are important for us to understand the nature of quantum gravity. Recently, by treating the cosmological constant as the pressure of the black hole thermal system\cite{Kastor:2009wy,Dolan:2010ha,Cvetic:2010jb,Dolan:2011xt}, the study of black hole thermodynamic has been extended into a more general case, in which the thermodynamical volume is defined as the conjugate variable of the pressure, and the mass of the black hole is considered as the enthalpy of the system. This proposal has inspired plenty of  interesting researches and remarkable phase structures of AdS black holes,  please see Ref. \cite{Kubiznak:2016qmn} and therein for  nice reviews.

In the framework of the extended thermodynamic of the black hole, the idea of defining a traditional heat engine via an AdS black hole  was proposed in \cite{Johnson:2014yja}. In details, the heat engine is realized by a circle in the pressure-volume phase space of the black hole. The input of heat ($Q_H$), the exhaust of heat ($Q_C$) and the mechanical work ($W$)  can be evaluated from the black hole system. Since the engine circle represents a process defined on the space of the  dual field theory living in one dimension lower, so that this kind of engine defined via an AdS black hole is named as holographic heat engine\cite{Johnson:2014yja}.  More remarkable properties of holographic heat engine were widely studied in black holes with Gauss-Bonnet correction\cite{Johnson:2015ekr}, in Born-Infeld corrected black hole\cite{Johnson:2015fva}, in rotational black hole\cite{Hennigar:2017apu,Johnson:2017ood}, in three dimensional black hole\cite{Mo:2017nhw}, and so on\cite{Liu:2017baz,Wei:2016hkm}.
%
%\begin{figure}[h]
%{\centering
%\includegraphics[width=3in]{heat_engine_flow.jpg}
 %  \caption{ The  heat engine flows.}   \label{fig:heatengine}}
%\end{figure}

However, until now the holographic heat engines  were mainly defined via the black holes dual to the boundary theory with translational symmetry, i.e. without momentum relaxation. It would be more realistic and  interesting to explore the heat engine defined via black holes with momentum relaxation, because our world is far from being ideal. Thus, the aim of this work is to study how the momentum relaxation of the black hole  affects the efficiency of the related holographic heat engine.

We will focus on the heat engine in the Einstein-Maxwell-Axions theory with negative cosmological constant, where the momentum relaxation is introduced by the linear massless axion fields\cite{Andrade:2013gsa}. We will mainly study the influence of momentum relaxation on the efficiency of the heat engine. Borrowing holography, it was found that the momentum relaxation brings  many novel properties in the dual condensed matter sectors\cite{Kim:2014bza,KimDNA,massless3,Fang:2015dia}, because it modifies the bulk geometry and breaks the translational symmetry of the boundary theory in a simple way. So we should expect that it will enrich the properties of the related heat engine due to the duality. Meanwhile, the study of the heat engine will conversely help to further understand the thermal systems with momentum relaxation and their holographic duality aspects.

The remaining of this paper is organized as follows. We briefly review the AdS black hole solution in Einstein-Maxwell-Axions theory, and study the extended thermodynamics in section \ref{sec:review}. Then in section \ref{sec:heat engine}, by giving two schemes of specified parameters, we will study how the efficiency of the  heat engine is affected by the momentum relaxation. Section \ref{sec:conclusion} contributes to our conclusions and discussions.

\section{Review of black hole in Einstein-Maxwell-Axions theory and the extended thermodynamics}
%%%%%%%%%%%%%%%%%%%%%%%%%%%%%%
\label{sec:review}

The Einstein-Maxwell-Axions gravity theory was proposed in \cite{Andrade:2013gsa}  by introducing $D-2$ massless scalar fields. The total action is given by
\begin{equation}
S=\frac{1}{16\pi }\int \! d^Dx \sqrt{-g} \left(R-2\Lambda -\frac{1}{4}F_{\mu\nu}F^{\mu\nu}-\frac{1}{2}\sum_{I=1}^{D-2}(\partial\psi_I)^2\right)\ ,
\label{eq:action}
\end{equation}
where  $D$ is the dimension of the spacetime, and the cosmological constant is
\begin{equation}
\Lambda=-\frac{(D-1)(D-2)}{2L^2}\ .
\label{eq:cosmocon}
\end{equation}

By setting the scalar fields to linearly depend on the $D-2$ dimensional spatial coordinates $x^a$, i.e., $\psi_I=\beta\delta_{Ia}x^a$\footnote{In general, the linear combination form of the scalar fields are $\psi_I=\beta_{Ia}x^a$. Then defining a constant $\beta^2\equiv \frac{1}{D-2}(\sum_{a=1}^{D-2}\sum_{I=1}^{D-2}\beta_{Ia}{\beta_{Ia}})$ with the coefficients satisfying the condition $\sum_{I=1}^{D-2}\beta_{Ia}{\beta_{Ib}}=\beta^2\delta_{ab}$, we will obtain the same black hole solution. Since there is rotational symmetry on the $x^a$ space, we can choose $\beta_{Ia}=\beta\delta_{Ia}$ without loss of generality.}
, one finds that the action admits the charged black hole solution
 \begin{eqnarray}\label{eq-metric}
&&ds^2=-f(r)dt^2+\frac{1}{f(r)}dr^2+r^2dx^a dx^a,~~~~~A=A_t(r) dt,~~\mathrm{with}\nonumber\\
&&f(r)=\frac{r^2}{L^2}-\frac{\beta^2}{2(D-3)}-\frac{m}{r^{D-3}}+\frac{q^2}{r^{2(D-3)}}, ~~~A_t=\sqrt{\frac{2(D-2)}{D-3}}\left(1-\frac{r_h^{D-3}}{r^{D-3}}\right)\frac{q}{r_h^{D-3}}
 \end{eqnarray}
where the index $a$ goes $a=1,2\cdot\cdot\cdot D-2 $, and the horizon $r_h$ satisfies $f(r_h)=0$. Note that the horizon has flat topology and the solution is valid only if $D\geqslant 4$.
It is worthwhile to point out that the scalar fields in the bulk source a spatially dependent field theory with momentum relaxation, which is dual to a homogeneous and isotropic black hole \eqref{eq-metric}. The linear coefficient $\beta$ of the scalar fields somehow can be considered to describe the strength of the momentum relaxation in the boundary theory\cite{Andrade:2013gsa}.

The mass and charge of the black hole are connected with the parameters $m$ and $q$ as
\begin{eqnarray}\label{eq-MQ}
M=\frac{(D-2)\mathcal{V}_{D-2}}{16\pi}m,~~~\mathrm{and}~~~Q=\frac{\sqrt{2(D-2)(D-3)}\mathcal{V}_{D-2}}{16\pi}q
\end{eqnarray}
where $\mathcal{V}_{D-2}$ is the volume of the $D-2$ dimensional flat space. The temperature of the black hole is
\begin{equation}\label{eq-T}
T=\frac{f'(r_h)}{4\pi}=\frac{1}{4\pi}\left(\frac{(D-1) r_h}{L^2}-\frac{\beta ^2}{2r_h}-\frac{(D-3)q^2}{r_h^{2D-5}}\right),
\end{equation}
and the entropy is
\begin{equation}\label{eq-S}
S=\frac{\mathcal{V}_{D-2}}{4}r_h^{D-2}.
\end{equation}

We will study the thermodynamic in the extended phase space. As proposed  in \cite{Kastor:2009wy,Dolan:2010ha,Cvetic:2010jb,Dolan:2011xt}, we assume the relation
\begin{equation}\label{eq-Pl}
P=-\frac{\Lambda}{8\pi}=\frac{(D-1)(D-2)}{16\pi L^2}.
\end{equation}
 Then the mass of the black hole can be rewritten as
\begin{equation}
M=\frac{(D-2)\mathcal{V}_{D-2}}{16\pi}\left(\frac{16 \pi  P r_h^{D-1}}{(D-1) (D-2)}+\frac{q^2}{r_h^{D-3}}-\frac{\beta^2 r_h^{D-3}}{2(D-3)}\right)
\end{equation}
which is defined as the enthapy $H$ of the system. Subsequently, the thermodynamical volume  is
\begin{equation}\label{eq-V}
V=\frac{\partial M}{\partial P}\big|_{S,Q}=\frac{\mathcal{V}_{D-2}}{D-1}r_h^{D-1},
\end{equation}
and the electric potential is \footnote{We can also calculate the electric potential $\Phi$, which is measured at infinity with respect to the horizon, via the method addressed in \cite{Caldarelli:1999xj,Cvetic:1999ne}, i.e., $\Phi=A_{t}|_{r\to \infty}-A_{t}|_{r\to r_h}=\sqrt{\frac{2(D-2)}{(D-3)}}\frac{q}{r_h^{D-3}}$. Here we have used the expression of $A_t$ in \eqref{eq-metric} and obviously we obtain vanishing static electric potential at the horizon.}
\begin{equation}
\Phi=\frac{\partial M}{\partial Q}\big|_{S,P}=\frac{16\pi Q}{(D-3)\mathcal{V}_{D-2}r_h^{D-3}}=\sqrt{\frac{2(D-2)}{(D-3)}}\frac{q}{r_h^{D-3}}
\end{equation}
where in the second equality, we used the relation in \eqref{eq-MQ}.

 \begin{figure}[h]
{\centering
\includegraphics[width=3in]{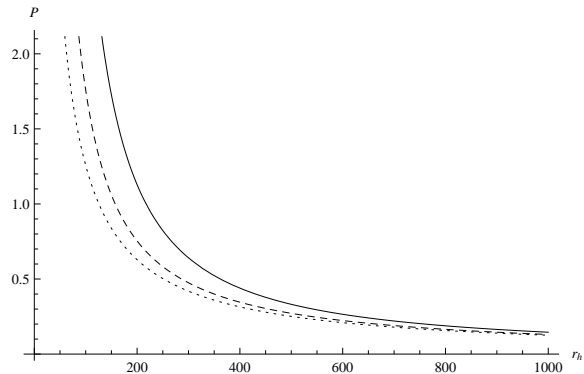}
   \caption{$P-r_h$ diagram for four dimensional black hole with momentum relaxation.  We fix $T=10$ and increase $\beta=0,100,200$ in the lines from the bottom to the top. In the plot, we set $q=1$ and $D=4$. }   \label{fig:p-rh}
}
\end{figure}

It is straightforward to verify that the first law of thermodynamics is
\begin{equation}\label{eq-dM}
dM=T dS+\Phi dQ+V dP+\varphi d\beta
\end{equation}
where $\varphi=\frac{\partial M}{\partial \beta}\big|_{S,P,Q}=-\frac{(D-2)\beta\mathcal{V}_{D-2}r_h^{D-3}}{16\pi(D-3)}$. And we obtain the Smarr relation for the black hole is
\begin{equation}\label{eq-Smarr}
(D-3)M=(D-2)TS+(D-3)\Phi Q-2PV.
\end{equation}
Note that in \eqref{eq-dM}, we see that $\beta$ is a variable in the law, however, in the above Smarr relation,
$\beta$ and its conjugation $\varphi$ do not have explicit contributions. This observation is reasonable because, after careful comparison, we find that it is consistent with the contribution of $c_2$ term in massive black hole to the extended thermodynamics \cite{Xu:2015rfa}.

We then substitute the pressure \eqref{eq-Pl} into the temperature \eqref{eq-T}, after which we can solve out the pressure as
\begin{equation}\label{eq-P}
P=\frac{(D-2)}{16\pi}\left(\frac{4\pi T}{r_h} +\frac{\beta^2}{2r_h^2}+\frac{(D-3)q^2}{r_h^{2D-4}}\right)\ .
\end{equation}
The above expression is also treated as the state equation.
From \eqref{eq-P}, we can figure out the $P-r_h$ or $P-V$ diagram with fixed temperature and momentum relaxation. The results are shown in figure \ref{fig:p-rh}. It is noted that here we are not interested in the existence of $P-V$ criticality in the extended phase space proposed in Refs.\cite{Chamblin:1999tk,Chamblin:1999hg,Kubiznak:2012wp}.

\section{Holographic heat engine via black hole with momentum relaxation}\label{sec:heat engine}
In this section, we intend  to define kind of heat engine via the black hole with momentum relaxation described in last section.

\subsection{The general engine efficiency  }
Before studying the efficiency of the engine,
we have to calculate the specific heat  $T\partial S/\partial T$, which can be computed from the expressions for temperature and the entropy. We treat both $T$ and $S$ as functions of the horizon $r_h$. Then from \eqref{eq-S}, we obtain
\begin{equation}
\frac{\partial S}{\partial r_h}=\frac{(D-2)\mathcal{V}_{D-2}}{4}r_h^{D-3}.
\end{equation}
Differentiation of the state equation \eqref{eq-T} gives us
\begin{equation}
dT=\frac{1}{4\pi}\left[  \left(\frac{16\pi P}{D-2}+\frac{\beta^2}{2r_h^2}+\frac{(D-3)(2D-5)q^2}{r_h^{2D-4}}\right)d r_h+\frac{16\pi r_h}{D-2}dP\right]
\end{equation}
from which we can reduce
\begin{equation}
\frac{\partial T}{\partial r_h}=\frac{\frac{16\pi P}{D-2}+\frac{\beta^2}{2r_h^2}+\frac{(D-3)(2D-5)q^2}{r_h^{2D-4}}}{4\pi-\frac{16\pi r_h}{D-2}\frac{\partial P}{\partial T}}.
\end{equation}
Subsequently, we have the general formula of the specific heat
 \begin{eqnarray}\label{eq-C}
 C&=&T\frac{\partial S}{\partial T}=T\frac{\left(\frac{\partial S}{\partial r_h}\right)}{\left(\frac{\partial T}{\partial r_h}\right)}\nonumber\\
 &=&\left(1-\frac{4 r_h}{D-2}\frac{\partial P}{\partial T}\right)\left(\frac{\frac{16\pi }{D-2} P\, r_h^{2D-4}-\frac{\beta^2 r_h^{2D-6}}{2}-(D-3)q^2 }{\frac{16\pi }{D-2} P\,r_h^{2D-4}+\frac{\beta^2 r_h^{2D-6}}{2}+{(D-3) (2D-5)q^2 }}\right)\frac{(D-2)\mathcal{V}_{D-2}}{4}r_h^{D-2}\ .
 \end{eqnarray}

From equation \eqref{eq-V},  it is obvious that constant volume means also constant $r_h$. And equation~(\ref{eq-P}) gives us
$(\partial P/\partial T)_V=(D-2)/4r_h$, so the specific heat at constant volume is zero,
 \begin{equation}
 C_V=T\frac{\partial S}{\partial T}\big|_V=0.
 \end{equation}
 While the specific heat at constant pressure, $C_P$, can be computed by setting $\partial P/\partial T=0$ in equation \eqref{eq-C}
  \begin{eqnarray}\label{eq-Cp}
 C_P&=&T\frac{\partial S}{\partial T}\big|_P=\left(\frac{\frac{16\pi }{D-2} P\, r_h^{2D-4}-\frac{\beta^2 r_h^{2D-6}}{2}-(D-3)q^2 }{\frac{16\pi }{D-2} P\,r_h^{2D-4}+\frac{\beta^2 r_h^{2D-6}}{2}+{(D-3) (2D-5)q^2 }}\right)\frac{(D-2)\mathcal{V}_{D-2}}{4}r_h^{D-2}.
 \end{eqnarray}
\begin{figure}[h]
{\centering
\includegraphics[width=3in]{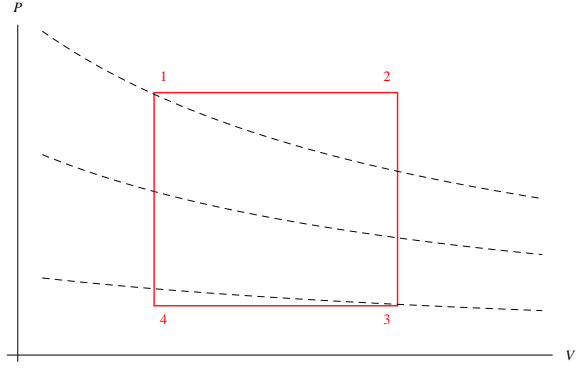}
   \caption{Cartoon of the engine.}   \label{fig-engine}}
\end{figure}

In order to define a heat engine, we will consider a rectangle cycle in the $P-V$ plane as it was done in the previous literatures\cite{Johnson:2015fva,Hennigar:2017apu,Johnson:2017ood,Mo:2017nhw,Liu:2017baz,Wei:2016hkm} . The rectangle consists of two isobars and two isochores as shown in figure \ref{fig-engine}. The vanishing of $C_V$ describes  the ``isochores'' while the heat flow along the isobars can be evaluated if $C_P$ is determined. Note that $1,2,3,4$ are four corners in the thermal flow cycle. In the following discussions, we will use the subscripts $1, 2, 3, 4$ to denote the relevant quantities evaluated at the related corners. Thus, in the cycle,  the work done by the engine is only along the isobar, i.e.,
\begin{equation}
W=(V_2-V_1)(P_1-P_4),
\label{eq-W}
\end{equation}
and the input of the heat is
\begin{equation}
\mathcal{Q}_H=\int_{T_1}^{T_2} C_P(P_1,T) dT\ .
 \label{eq-Q}
\end{equation}
So the engine efficiency is $\eta=W/\mathcal{Q}_H$.
Note that in $P-V$ plane, the isotherms at temperatures $T_h$ and $T_l$ with $T_h>T_l$ give the Carnot efficiency $\eta_C=1-T_l/T_h$. And for our engine, it is
\begin{equation}
\eta_C=1-\frac{T_l}{T_h}=1-\frac{T_4}{T_2}.
\end{equation}

In what follows, we will study $\eta$ for the heat engine and focus on the effect of momentum relaxation. We shall first work in four dimensional theory, and then only show the results for five dimensional case in the appendix \ref{appendix}. We will set $q=0.1$ and the volume $\mathcal{V}_{2}=1$ without loss of generality.

\subsection{Engine efficiency in large temperature limit}
In order to evaluate the efficiency,  we have to calculate the expressions \eqref{eq-W} and \eqref{eq-Q}. Usually, it is difficult to do the exact integration in  \eqref{eq-Q}. So we will consider the large temperature limit, i.e., $T \gg \beta,q$, which means that $1/T$ can be treated as a small quantity\footnote{Note that the study with $\beta=0$ goes back to that in Reissner-Nordstr$\ddot{o}$m black hole but with spherical horizon in \cite{Johnson:2014yja}. Here  $\beta$ can be arbitrary real constants in the black hole solution. However, in our study, $\beta$  always appears in the form $\beta^2$, so we will focus on real positive $\beta$ to see its effect without loss of generality. Moreover, we  will study the heat engine efficiency in large temperature limit with $T \gg \beta,q$. It is also interesting to study the effect of more general $\beta$. }.
We first solve out $r_h$ in term of large $T$ from \eqref{eq-T},
\begin{eqnarray}
r_h=\frac{T}{2P}+\frac{\beta^2}{8\pi T}+\frac{P(32\pi P q^2-\beta^4)}{32\pi^2 T^3}+\cdots\ .
\end{eqnarray}
Then from \eqref{eq-V} and \eqref{eq-Cp}, we can obtain
\begin{eqnarray}
&&V=\frac{1}{3}r_h^3= \frac{T^3}{24 P^3}+\frac{\beta^2T}{32\pi P^2}+\frac{q^2}{4\pi T}+\cdots\ ,\nonumber \\
&&C_P=\frac{T^2}{8P^2}+\frac{-64\pi Pq^2+\beta^2}{128\pi^2T^2}+\frac{P\beta^2(80\pi Pq^2-\beta^4)}{64\pi^3 T^4}+\cdots,
\label{eq:4dexpansions}
\end{eqnarray}
respectively.

Consequently,  the  efficiency of the engine can be obtained
\begin{eqnarray}
\eta=\frac{W}{\mathcal{Q}_H}&=&\left(1-\frac{P_4}{P_1}\right)\times\frac{P_1(V_2-V_1)}{\int_{T_1}^{T_2}C_P(P_1,T)dT}\nonumber\\
&\simeq&\left(1-\frac{P_4}{P_1}\right)\times\left(\frac{\frac{1}{24P_1^2}(T_2^3-T_1^3)+\frac{\beta^2}{32\pi P_1}(T_2-T_1)}{\frac{1}{24P_1^2}(T_2^3-T_1^3)}+\cdots\nonumber\right)\\
&\simeq&\left(1-\frac{P_4}{P_1}\right)\left(1+\frac{3\beta^2}{4\pi}\frac{P_1}{T_2^2+T_2T_1+T_1^2}+\cdots\right)
\label{eq-efficiency1}
\end{eqnarray}
where the volumes $V_1$ and $V_2$ correspond to the volume for $T_1$ and $T_2$ at $P=P_1=P_2$, respectively.

We shall disclose how the momentum relaxation $\beta$ affects the efficiency of our heat engine by evaluating the ratio $\eta/\eta_C$ and $\eta/\eta_0$, where $\eta_0$ is the efficiency at $\beta=0$.  We will consider two schemes of specified parameters in the engine circle as pioneerly  addressed in
 \cite{Johnson:2015ekr}.  Note that in order to make sure that the dual heat engine is thermodynamically physical, we have to set the proper parameters to satisfy the efficiency lower than unit.

\begin{figure}[h]
{\centering
\includegraphics[width=5in]{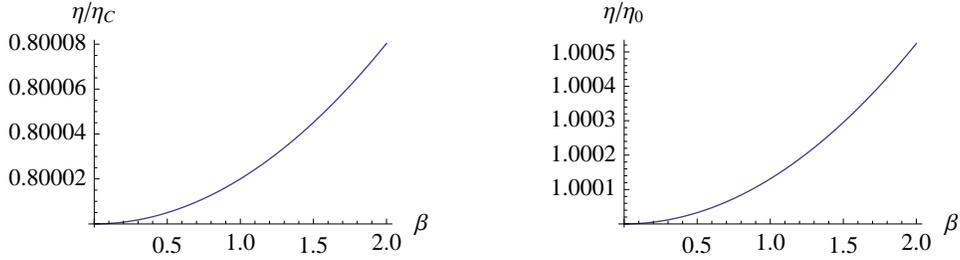}
   \caption{Results in the fist scheme with $D=4$. Left: the ratio $\eta/\eta_C$. Right: the ratio $\eta/\eta_0$. We have chosen the specific parameters of the cycle: $P_1 = 5, P_4 = 3, T_1 = 50, T_2 = 60$. }   \label{fig-eta1}}
\end{figure}
Firstly, we choose the engine cycle with specified $(T_1,T_2,P_1,P_4)$, i.e. we set $P_1 = 5, P_4 = 3, T_1 = 50$ and $T_2 = 60$ for the engine in figure \ref{fig-engine}. Recalling the state equation \eqref{eq-P}, we can first calculate $V_1$ through the given $(T_1,P_1)$, then compute  the temperature of heat source $T_l=T_4$ through the known $(P_4, V_4=V_1)$ . It is obvious that in this scheme $T_4$ decreases as $\beta$ increases, such that the Carnot efficiency $\eta_C=1-T_4/T_2$ is enhanced by $\beta$. On the other hand, we see from \eqref{eq-efficiency1} that the efficiency $\eta$ also increases when we increase the momentum relaxation. So, it is not direct to say how the ratio $\eta/\eta_C$ changes as $\beta$. We show the results in figure \ref{fig-eta1}. The left plot shows that the efficiency is lower than the Carnot efficiency and the ratio $\eta/\eta_C$ increases when momentum relaxation becomes stronger. While,  the right plot shows that the efficiency is higher for  bigger $\beta$, which is explicit from equation \eqref {eq-efficiency1} with fixed $(T_1,T_2,P_1,P_4)$.

\begin{figure}[h]
{\centering
\includegraphics[width=5in]{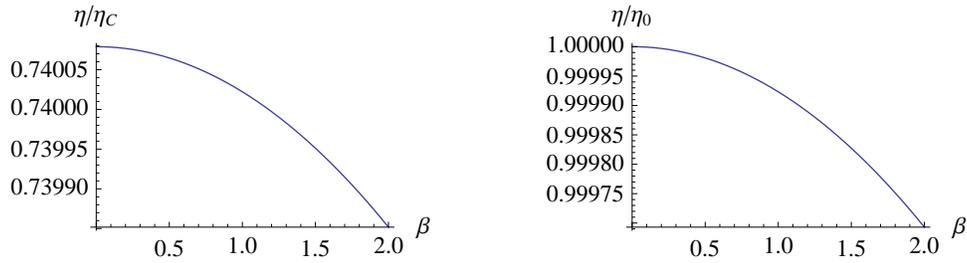}
   \caption{Results in the second scheme with $D=4$. Left: the ratio $\eta/\eta_C$. Right: the ratio $\eta/\eta_0$. We have chosen the specific parameters of the cycle: $T_2 = 60, T_4 = 30, V_2 = 3000, V_4 = 1500$. We work in $D=4$.}   \label{fig-eta2}}
\end{figure}
Let us turn to study the engine efficiency with specified $(T_2,T_4,V_2,V_4)$. In this scheme, the Carnot efficiency will not change with $\beta$. Similarly, we recall the state equation again. We can determine $P_1=P_2$ and $P_4$ via $(T_2,V_2)$ and  $(T_4,V_4)$, respectively. Then we calculate $T_1$ via $(V_1=V_4,P_1)$. Our results of the efficiency in this scheme are shown in figure \ref{fig-eta2}. The tendencies of $\eta/\eta_C$ and $\eta/\eta_0$ are both suppressed by larger $\beta$, which are in contrast to those occur in the first scheme.

The results of the efficiency for $D=5$ are shown in appendix \ref{appendix}. It is obvious that for the second scheme, the behaviors of $\eta/\eta_C$ and $\eta/\eta_0$ depending on $\beta$ are qualitatively the same as those in $D=4$ case. However, for the first scheme, the ratio $\eta/\eta_C$ deceases as $\beta$ increases, which is different from the behavior in $D=4$ dimensional  theory.

\section{Conclusion and discussion}\label{sec:conclusion}
In this paper, we extended the study of the heat engine by defining it via a simple black hole with momentum relaxation $\beta$. We especially studied the effect of momentum relaxation on the efficiency of heat engine in large temperature limit. Momentum relaxation introduced by linear massless axion fields has been widely studied in holographic condensed matter topics, because it breaks the translational symmetry of the dual field theory and gives finite conductivity in a simple way.
$\beta$ in this model modifies the state equation in the extended thermodynamics of the black hole, so that it  has print on the heat capacity. Thus, the study of heat engine affected by momentum relaxation may bring insights to both the dual boundary theory and the thermodynamics of the black hole.

In two different schemes with specified parameters, we evaluated the efficiency $\eta$ comparing to both the Carnot efficiency $\eta_C$ and the efficiency with $\beta=0$, denoted $\eta_0$. In the first scheme  with given $(T_1,T_2,P_1,P_4)$, as $\beta$ becomes stronger, both $\eta/\eta_C$  and $\eta/\eta_0$ increases. In the second scheme with given $(T_2,T_4,V_2,V_4)$, both $\eta/\eta_C$ and $\eta/\eta_0$ decrease as $\beta$ increases,  which are different from those happen in the first scheme. The effects of momentum relaxation on the efficiency for the second scheme in our model are opposite  to the effects of Born-Infeld correction in the Maxwell field in four dimensional case addressed in \cite{Johnson:2015fva}. While for the first scheme, the behavior $\eta/\eta_C$ obeys the same rule but $\eta/\eta_0$ behaves in an opposite way. Moreover, comparing to the effect of Gauss-Bonnet correction\cite{Johnson:2015ekr}, here the effect of $\beta$ also qualitatively depends on the dimension of the gravity theory. In five dimensional thoery,
$\eta/\eta_0$ increases in both theories while $\eta/\eta_C$ behaves oppositely in the first scheme, but in the second scheme,
both $\eta/\eta_C$ and $\eta/\eta_0$ decrease as Gauss-Bonnet coupling or momentum relaxation increases. In our previous paper
\cite{Kuang:2017cgt}, we found the black hole solution in arbitrary dimensional charged Gauss-Bonnet-Maxwell-Axions theory, so it would be very interesting to study the mixed  effects of Gauss-Bonnet coupling and momentum relaxation, from which we may see the enhance/competition of increasing the efficiency of the holographic heat engine. We shall report the results in the near future.

The gravity theory we chose to study the heat engine is a simple theory dual to field theory with momentum relaxation. A more general homogeneous theory without translational symmetry is the massive gravity proposed in \cite{Vegh:2013sk,Blake:2013owa}. The extended thermodynamics of the massive black holes  has been investigated in \cite{Xu:2015rfa,Hendi:2017fxp}. Very recently, the efficiency of the heat engine via black holes in massive gravity was addressed in \cite{Mo:2017nes,Hendi:2017bys} where the authors studied the effects of more general bulk parameters on the  efficiency of the heat engine. Note that both our studies of the effects of momentum relaxation on the heat engine focused on four or five dimensional background.  It would be interesting to study the heat engine via three dimensional black holes with momentum relaxation, for example the BTZ massive black hole constructed in \cite{Hendi:2016pvx}.

\section*{Acknowledgements}
This work is supported by the Natural Science
Foundation of China under Grant Nos. 11505116 and 11705161. L. Q. Fang is also supported by Natural Science Foundation of Jiangxi Province under Grant No. 20171BAB211013.
X. M. Kuang is also supported by Natural Science Foundation of Jiangsu Province under Grant No. BK20170481.
\appendix
\section{Results in five dimensional theory}\label{appendix}
\begin{figure}[h]
{\centering
\includegraphics[width=5in]{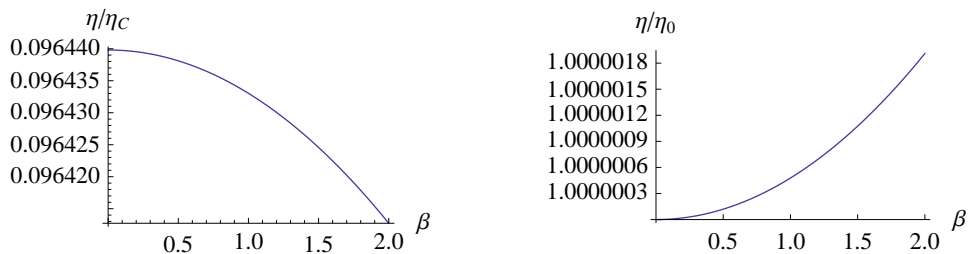}
   \caption{Results in the fist scheme with $D=5$. Left: the ratio $\eta/\eta_C$. Right: the ratio $\eta/\eta_0$. We have chosen the specific parameters of the cycle: $P_1 = 5, P_4 = 3, T_1 = 50, T_2 = 60$.}   \label{fig-eta3}}
\end{figure}

\begin{figure}[h]
{\centering
\includegraphics[width=5in]{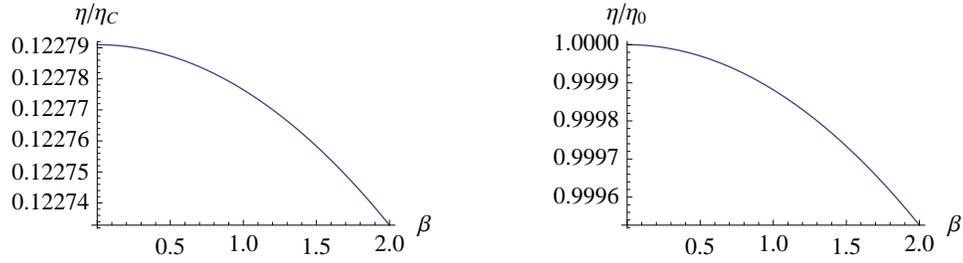}
   \caption{Results in the second scheme with $D=5$. Left: the ratio $\eta/\eta_C$. Right: the ratio $\eta/\eta_0$. We have chosen the specific parameters of the cycle: $T_2 = 60, T_4 = 30, V_2 = 3000, V_4 = 1500$. }   \label{fig-eta4}}
\end{figure}

In five dimensional theory, we obtain in large temperature limit that
\begin{eqnarray}
r_h&=&\frac{3T}{4 P}+\frac{\beta ^2}{8 \pi T }-\frac{\beta ^4 P}{48 \pi ^2 T^3}+\cdot\cdot\cdot\nonumber\\
V&=&\frac{9T^3}{64 P^3}+\frac{9 \beta ^2T}{128 \pi  P^2}-\frac{5 \beta ^6}{1536 \pi ^3T^3}+\cdot\cdot\cdot\nonumber\\
C_P&=&\frac{81 T^3}{256 P^3}+\frac{27 \beta ^2T}{512 \pi  P^2}+ \left(-\frac{21 \beta ^6}{2048 \pi ^3}-\frac{4 P^2 q^2}{\pi }\right)\frac{1}{T^3}+\cdot\cdot\cdot
\end{eqnarray}
and the efficiency is
\begin{eqnarray}
\eta=\left(1-\frac{P_4}{P_1}\right)\left[\frac{\frac{9}{64P_1^2}(T_2^3-T_1^3)+\frac{9\beta^2}{128\pi P_1}(T_2-T_1)+\cdot\cdot\cdot} {\frac{81}{1024P_1^3}(T_2^4-T_1^4)+\frac{27\beta^2}{1024\pi P_1^2}(T_2^2-T_1^2)+\cdot\cdot\cdot}\right].
\end{eqnarray}
The behaviors of $\eta/\eta_C$ and $\eta/\eta_0$ are shown in figure \ref{fig-eta3} and figure \ref{fig-eta4}, which have some different qualitative features from the results in $D=4$ dimensional case. The detailed analysis of the differences can be seen in the last paragraph of section \ref{sec:heat engine}.

%\bibliographystyle{../preprint}
%\bibliography{../Bib/QuantGra.bib}

\begin{thebibliography}{10}

%\cite{Kastor:2009wy}
\bibitem{Kastor:2009wy}
  D.~Kastor, S.~Ray and J.~Traschen,
  ``Enthalpy and the Mechanics of AdS Black Holes,''
  Class.\ Quant.\ Grav.\  {\bf 26}, 195011 (2009)
  %doi:10.1088/0264-9381/26/19/195011
  [arXiv:0904.2765 [hep-th]].
  %%CITATION = doi:10.1088/0264-9381/26/19/195011;%%
  %288 citations counted in INSPIRE as of 29 May 2017

%\cite{Dolan:2010ha}
\bibitem{Dolan:2010ha}
  B.~P.~Dolan,
  ``The cosmological constant and the black hole equation of state,''
  Class.\ Quant.\ Grav.\  {\bf 28}, 125020 (2011)
  %doi:10.1088/0264-9381/28/12/125020
  [arXiv:1008.5023 [gr-qc]].
  %%CITATION = doi:10.1088/0264-9381/28/12/125020;%%
  %176 citations counted in INSPIRE as of 29 May 2017

  %\cite{Cvetic:2010jb}
\bibitem{Cvetic:2010jb}
  M.~Cvetic, G.~W.~Gibbons, D.~Kubiznak and C.~N.~Pope,
  ``Black Hole Enthalpy and an Entropy Inequality for the Thermodynamic Volume,''
  Phys.\ Rev.\ D {\bf 84}, 024037 (2011)
  %doi:10.1103/PhysRevD.84.024037
  [arXiv:1012.2888 [hep-th]].
  %%CITATION = doi:10.1103/PhysRevD.84.024037;%%
  %217 citations counted in INSPIRE as of 29 May 2017

 %\cite{Dolan:2011xt}
\bibitem{Dolan:2011xt}
  B.~P.~Dolan,
  ``Pressure and volume in the first law of black hole thermodynamics,''
  Class.\ Quant.\ Grav.\  {\bf 28}, 235017 (2011)
  %doi:10.1088/0264-9381/28/23/235017
  [arXiv:1106.6260 [gr-qc]].
  %%CITATION = doi:10.1088/0264-9381/28/23/235017;%%
  %175 citations counted in INSPIRE as of 29 May 2017

  %\cite{Kubiznak:2016qmn}
\bibitem{Kubiznak:2016qmn}
  D.~Kubiznak, R.~B.~Mann and M.~Teo,
  ``Black hole chemistry: thermodynamics with Lambda,''
  Class.\ Quant.\ Grav.\  {\bf 34}, no. 6, 063001 (2017)
  %doi:10.1088/1361-6382/aa5c69
  [arXiv:1608.06147 [hep-th]].
  %%CITATION = doi:10.1088/1361-6382/aa5c69;%%
  %36 citations counted in INSPIRE as of 29 May 2017

  %\cite{Johnson:2014yja}
\bibitem{Johnson:2014yja}
  C.~V.~Johnson,
  ``Holographic Heat Engines,''
  Class.\ Quant.\ Grav.\  {\bf 31}, 205002 (2014)
  %doi:10.1088/0264-9381/31/20/205002
  [arXiv:1404.5982 [hep-th]].
  %%CITATION = doi:10.1088/0264-9381/31/20/205002;%%
  %75 citations counted in INSPIRE as of 29 May 2017

%\cite{Johnson:2015ekr}
\bibitem{Johnson:2015ekr}
  C.~V.~Johnson,
  ``Gauss-Bonnet black holes and holographic heat engines beyond large $N$,''
  Class.\ Quant.\ Grav.\  {\bf 33}, no. 21, 215009 (2016)
  %doi:10.1088/0264-9381/33/21/215009
  [arXiv:1511.08782 [hep-th]].
  %%CITATION = doi:10.1088/0264-9381/33/21/215009;%%
  %18 citations counted in INSPIRE as of 29 May 2017

  %\cite{Johnson:2015fva}
\bibitem{Johnson:2015fva}
  C.~V.~Johnson,
  ``Born-Infeld AdS black holes as heat engines,''
  Class.\ Quant.\ Grav.\  {\bf 33}, no. 13, 135001 (2016)
  %doi:10.1088/0264-9381/33/13/135001
  [arXiv:1512.01746 [hep-th]].
  %%CITATION = doi:10.1088/0264-9381/33/13/135001;%%
  %11 citations counted in INSPIRE as of 29 May 2017

  %\cite{Hennigar:2017apu}
\bibitem{Hennigar:2017apu}
  R.~A.~Hennigar, F.~McCarthy, A.~Ballon and R.~B.~Mann,
  ``Holographic heat engines: general considerations and rotating black holes,''
  arXiv:1704.02314 [hep-th].
  %%CITATION = ARXIV:1704.02314;%%
  %2 citations counted in INSPIRE as of 29 May 2017

%\cite{Johnson:2017ood}
\bibitem{Johnson:2017ood}
  C.~V.~Johnson,
  ``Taub-Bolt Heat Engines,''
  arXiv:1705.04855 [hep-th].
  %%CITATION = ARXIV:1705.04855;%%

%\cite{Mo:2017nhw}
\bibitem{Mo:2017nhw}
  J.~X.~Mo, F.~Liang and G.~Q.~Li,
  ``Heat engine in the three-dimensional spacetime,''
  JHEP {\bf 1703}, 010 (2017)
  %doi:10.1007/JHEP03(2017)010
  [arXiv:1701.00883 [gr-qc]].
  %%CITATION = doi:10.1007/JHEP03(2017)010;%%

%\cite{Liu:2017baz}
\bibitem{Liu:2017baz}
  H.~Liu and X.~H.~Meng,
  ``Effects of dark energy on the efficiency of charged AdS black holes as heat engine,''
  arXiv:1704.04363 [hep-th].
  %%CITATION = ARXIV:1704.04363;%%
  %1 citations counted in INSPIRE as of 29 May 2017

%\cite{Wei:2016hkm}
\bibitem{Wei:2016hkm}
  S.~W.~Wei and Y.~X.~Liu,
  %``Implementing black hole as efficient power plant,''
  arXiv:1605.04629 [gr-qc].
  %%CITATION = ARXIV:1605.04629;%%
  %5 citations counted in INSPIRE as of 29 May 2017

%\cite{Andrade:2013gsa}
\bibitem{Andrade:2013gsa}
  T.~Andrade and B.~Withers,
  ``A simple holographic model of momentum relaxation,''
  JHEP {\bf 1405}, 101 (2014)
  %doi:10.1007/JHEP05(2014)101
  [arXiv:1311.5157 [hep-th]].
  %%CITATION = doi:10.1007/JHEP05(2014)101;%%
  %143 citations counted in INSPIRE as of 29 May 2017

%\cite{Kim:2014bza}
\bibitem{Kim:2014bza}
  K.~Y.~Kim, K.~K.~Kim, Y.~Seo and S.~J.~Sin,
  ``Coherent/incoherent metal transition in a holographic model,''
  JHEP {\bf 1412}, 170 (2014)
  %doi:10.1007/JHEP12(2014)170
  [arXiv:1409.8346 [hep-th]].
  %%CITATION = doi:10.1007/JHEP12(2014)170;%%
  %50 citations counted in INSPIRE as of 29 May 2017

\bibitem{KimDNA}
  K.~Y.~Kim, K.~K.~Kim and M.~Park,
  A Simple Holographic Superconductor with Momentum Relaxation,
JHEP {\bf 1504}, 152 (2015).
[arXiv:1501.00446 [hep-th]].
%%CITATION = arXiv:1501.00446%%
\bibitem{massless3} B. Gout$\acute{\textmd{e}}$raux, Charge transport in holography with momentum dissipation, JHEP 1404
(2014) 181, [arXiv:1401.5436].

%\cite{Fang:2015dia}
\bibitem{Fang:2015dia}
  L.~Q.~Fang, X.~M.~Kuang, B.~Wang and J.~P.~Wu,
  ``Fermionic phase transition induced by the effective impurity in holography,''
  JHEP {\bf 1511}, 134 (2015)
  %doi:10.1007/JHEP11(2015)134
  [arXiv:1507.03121 [hep-th]].
  %%CITATION = doi:10.1007/JHEP11(2015)134;%%
  %5 citations counted in INSPIRE as of 29 May 2017

%\cite{Caldarelli:1999xj}
\bibitem{Caldarelli:1999xj}
  M.~M.~Caldarelli, G.~Cognola and D.~Klemm,
  ``Thermodynamics of Kerr-Newman-AdS black holes and conformal field theories,''
  Class.\ Quant.\ Grav.\  {\bf 17}, 399 (2000)
  %doi:10.1088/0264-9381/17/2/310
  [hep-th/9908022].
%\cite{Cvetic:1999ne}
\bibitem{Cvetic:1999ne}
  M.~Cvetic and S.~S.~Gubser,
  ``Phases of R charged black holes, spinning branes and strongly coupled gauge theories,''
  JHEP {\bf 9904}, 024 (1999)
  %doi:10.1088/1126-6708/1999/04/024
  [hep-th/9902195].
  %%CITATION = doi:10.1088/1126-6708/1999/04/024;%%

\bibitem{Xu:2015rfa}
  J.~Xu, L.~M.~Cao and Y.~P.~Hu,
  ``P-V criticality in the extended phase space of black holes in massive gravity,''
  Phys.\ Rev.\ D {\bf 91}, no. 12, 124033 (2015)
  %doi:10.1103/PhysRevD.91.124033
  [arXiv:1506.03578 [gr-qc]].
  %%CITATION = doi:10.1103/PhysRevD.91.124033;%%

%\cite{Chamblin:1999tk}
\bibitem{Chamblin:1999tk}
  A.~Chamblin, R.~Emparan, C.~V.~Johnson and R.~C.~Myers,
  ``Charged AdS black holes and catastrophic holography,''
  Phys.\ Rev.\ D {\bf 60}, 064018 (1999)
  %doi:10.1103/PhysRevD.60.064018
  [hep-th/9902170].
  %%CITATION = doi:10.1103/PhysRevD.60.064018;%%
  %658 citations counted in INSPIRE as of 29 May 2017

%\cite{Chamblin:1999hg}
\bibitem{Chamblin:1999hg}
  A.~Chamblin, R.~Emparan, C.~V.~Johnson and R.~C.~Myers,
  ``Holography, thermodynamics and fluctuations of charged AdS black holes,''
  Phys.\ Rev.\ D {\bf 60}, 104026 (1999)
  %doi:10.1103/PhysRevD.60.104026
  [hep-th/9904197].
  %%CITATION = doi:10.1103/PhysRevD.60.104026;%%
  %358 citations counted in INSPIRE as of 29 May 2017

%\cite{Kubiznak:2012wp}
\bibitem{Kubiznak:2012wp}
  D.~Kubiznak and R.~B.~Mann,
  ``P-V criticality of charged AdS black holes,''
  JHEP {\bf 1207}, 033 (2012)
  %doi:10.1007/JHEP07(2012)033
  [arXiv:1205.0559 [hep-th]].
  %%CITATION = doi:10.1007/JHEP07(2012)033;%%
  %265 citations counted in INSPIRE as of 29 May 2017

%\cite{Kuang:2017cgt}
\bibitem{Kuang:2017cgt}
  X.~M.~Kuang and J.~P.~Wu,
  ``Thermal transport and quasi-normal modes in Gauss?Bonnet-axions theory,''
  Phys.\ Lett.\ B {\bf 770}, 117 (2017)
  %doi:10.1016/j.physletb.2017.04.045
  [arXiv:1702.01490 [hep-th]].
  %%CITATION = doi:10.1016/j.physletb.2017.04.045;%%

%\cite{Vegh:2013sk}
\bibitem{Vegh:2013sk}
  D.~Vegh,
  ``Holography without translational symmetry,''
  arXiv:1301.0537 [hep-th].
  %%CITATION = ARXIV:1301.0537;%%
%\cite{Blake:2013owa}
\bibitem{Blake:2013owa}
  M.~Blake, D.~Tong and D.~Vegh,
  ``Holographic Lattices Give the Graviton an Effective Mass,''
  Phys.\ Rev.\ Lett.\  {\bf 112}, no. 7, 071602 (2014)
  %doi:10.1103/PhysRevLett.112.071602
  [arXiv:1310.3832 [hep-th]].
%\cite{Xu:2015rfa}

%\cite{Hendi:2017fxp}
\bibitem{Hendi:2017fxp}
  S.~H.~Hendi, R.~B.~Mann, S.~Panahiyan and B.~Eslam Panah,
  ``Van der Waals like behavior of topological AdS black holes in massive gravity,''
  Phys.\ Rev.\ D {\bf 95}, no. 2, 021501 (2017)
  %doi:10.1103/PhysRevD.95.021501
  [arXiv:1702.00432 [gr-qc]].
  %%CITATION = doi:10.1103/PhysRevD.95.021501;%%

%\cite{Mo:2017nes}
\bibitem{Mo:2017nes}
  J.~X.~Mo and G.~Q.~Li,
  ``Holographic Heat engine within the framework of massive gravity,''
  arXiv:1707.01235 [gr-qc].
%\cite{Hendi:2017bys}
\bibitem{Hendi:2017bys}
  S.~H.~Hendi, B.~Eslam Panah, S.~Panahiyan, H.~Liu and X.-H.~Meng,
  ``Black holes in massive gravity as heat engines,''
  arXiv:1707.02231 [hep-th].

%\cite{Hendi:2016pvx}
\bibitem{Hendi:2016pvx}
  S.~H.~Hendi, B.~Eslam Panah and S.~Panahiyan,
  ``Massive charged BTZ black holes in asymptotically (a)dS spacetimes,''
  JHEP {\bf 1605}, 029 (2016)
  %doi:10.1007/JHEP05(2016)029
  [arXiv:1604.00370 [hep-th]].
  %%CITATION = doi:10.1007/JHEP05(2016)029;%%

\end{thebibliography}

\end{document}